\documentstyle[aps,pre,twocolumn,floats]{revtex}

\begin{document}
\input{epsf.tex}

\draft
\epsfclipon

\date{\today}

\title{Formation of disclination lines near a free nematic interface}
\author{Jordi Ign\'es-Mullol, Jean Baudry, Lubor Lejcek%
\thanks{Permanent address: Institute of Physics, Na Slovance 2,%
18040 Prage, Czech Republic.}, and Patrick Oswald%
\thanks{Electronic address: oswald@physique.ens-lyon.fr.}}

\address{
\'Ecole Normale Sup\'erieure de Lyon, Laboratoire de Physique , 46 All\'ee 
d'Italie, 69364 Lyon Cedex 07, France}
\maketitle
\begin{abstract}
We have studied the nucleation and the physical properties 
of a -1/2 wedge disclination line near the
free surface of a confined nematic liquid crystal. The position of the 
disclination line has been related to the material parameters (elastic
constants, 
anchoring energy and favored anchoring angle of the molecules at the free
surface). 
The use of a planar model for the structure of the director field (whose
predictions have been contrasted to those of a fully three-dimensional
model) has allowed us to relate the experimentally
observed position of the disclination line
to the relevant properties of the liquid crystals. In particular, we have
been able to observe the collapse of the disclination line due to a 
temperature-induced anchoring angle transition, which has allowed us
to rule out the presence of a real disclination line near the nematic/isotropic
front in directional growth experiments.
\end{abstract}
\pacs {61.30.Jf,61.30.Gd}

\section{Introduction}
\label{sec:introduction}
Disclination lines \cite{Frank58} are characteristic defects of uniaxial 
nematic phases. These lines break the orientational order of the director 
field. They can easily form, for instance by stirring a nematic drop 
deposited on a glass plate. Observation in the microscope reveals a 
{\sl threaded texture} composed of thin and thick threads which are 
primarily $\pm 1/2$ and $\pm 1$ wedge disclination lines. 
The former have singular cores whereas the latter form continuous 
configurations since the director can escape along the disclination axis 
(the so-called {\sl escape in the third dimension} \cite{Dzyaloshinskii73}).
These lines cost an energy per unit length of the order of K 
(the relevant Frank constant), so they tend to spontaneously collapse when 
the flow vanishes. They can even all disappear at the end if the  
constraints at the boundaries and in the bulk are negligible (no external 
field). By contrast, some of them may subsist if the boundary conditions 
impose topological constraints which are incompatible with a continuous 
director field. Such a phenomenom occurs when the liquid crystal is 
confined inside a sphere with strong homeotropic anchoring (nematic
 molecules anchored parallel to the local normal of the surface). 
In this case, a point defect occurs at the center of the sphere 
(hedgehog configuration) \cite{Candau73,Volovik83}. 
This configuration is stable if the anchoring energy per unit surface area, 
$W$, (see below for a more detailed definition of this quantity) 
is strong enough to prevent the homeotropic anchoring to be broken. 
More precisely, a defect will appear if the anchoring penetration length 
$K/W$ \cite{Ryschenkow76} is smaller than the sphere radius $R$. 
Such configurations can be obtained by dispersing a nematic phase in a 
non miscible isotropic liquid \cite{Drzaic88}. 
They are also found in polymer dispersed liquid crystals (PDLC) 
\cite{Doane90}. 
These materials, of great practical interest, are obtained by a 
polymerization-induced phase separation with an average droplet radius 
typically ranging between 0.1 $\mu$m and 10 $\mu$m. Another way to nucleate a 
disclination line is to confine the nematic liquid crystal inside a 
capillary tube with a strong homeotropic anchoring. If the tube is circular, 
the line is a +1 disclination \cite{Cladis72}. 
On the other hand, two +1/2 disclinations must occur near both sides of a 
flat rectangular capillary tube \cite{Mihailovic88}. In all these examples 
the nematic phase forms a simply connected medium. Conversely, it is
possible to 
disperse a non-miscible liquid into a nematic liquid crystal and form this 
way an emulsion. This problem has recently attracted attention \cite{Poulin97}
because the elastic distorsions produced by the point defects that nucleate 
near each droplet create long range dipolar forces between the droplets.

In this article, we have chosen to study the nucleation and the physical 
properties of the -1/2 wedge disclination line that forms at the edge of a 
nematic sample confined between two parallel glass plates treated chemically 
to induce a homeotropic anchoring. At the edge, the nematic sample is in 
contact with an isotropic fluid. The interface between the nematic phase and 
the surrounding isotropic fluid is generally curved across the sample 
thickness because the two phases wet the glass plates differently.
The first experimentally addressed example was that of the nematic interface 
in contact with its own melt. Such a situation is achieved in directional 
solidification when a homeotopic nematic sample is placed in a temperature 
gradient so that the nematic/isotropic phase transition temperature lies 
somewhere in the sample. Previous experiments in directional solidification 
have shown that a disclination line can detach from the interface in the 
form of triangular domains \cite{Oswald87}. 
To explain this observation, it was suggested 
that a disclination line lies behind the meniscus separating the two phases 
and that a dust particle crossing the line could detach it from the 
meniscus, leading to the observed triangles. This interpretation was 
reinforced by the fact that energy considerations alone suggest this line 
can exist near the meniscus. Another reason to suppose its existence was 
that the wavelength measured at the onset of the Mullins-Sekerka instability  
was five times larger than the wavelength predicted by the linear stability 
analysis \cite{Bechhoefer89,Figueiredo93,Figueiredo96},
suggesting a possible renormalization of the apparent surface 
tension due to the disclination line and to the associated elastic effects 
\cite{Misbah95,Bechhoefer95} . 
Unfortunately the line was not visible 
in the microscope in the thin samples used in directional growth.  
For this reason we tried to see the line by increasing the sample thickness 
(its distance to the meniscus should essentially scale with the sample 
thickness). Surprisingly, even though we increased
the sample thickness up to 400 $\mu$m, 
we never saw the line and concluded that it might be virtual.
        
Prompted by this failure, we attempted to observe a disclination line at a 
free nematic interface (i.e., in contact with air), and we discovered that the 
line may or may not exist in that system, depending on both the strength of 
the anchoring energy and, much more important, on the value of the
preferred tilt 
angle of the molecules at the free interface.

This article is structured as follows: in section (\ref{sec:experiment}) we
detail the basic experimental setup and procedure, together with the main 
phenomenology involved in the formation of a disclination line near a free
interface in a confined nematic phase. 
In section (\ref{sec:model}) we outline a planar model
for the director field in the presence of a disclination line which we use
to do a numerical study. In section
(\ref{sec:model_3D}) a more realistic, three-dimensional structure
of the director field is considered and we show it is a good
approximation to use the planar model to calculate the position of the
disclination line. Finally, in section (\ref{sec:anchoring_transition}) we
consider
the effect of the anchoring angle on the disclination line, and use our planar
model to study a temperature-induced anchoring angle transition.

\section{Experiment}
\label{sec:experiment}
We have prepared thin nematic (N) liquid crystal
samples confined between two parallel glass plates.
In order to favor a homeotropic anchoring of the liquid crystal on the 
glass plates, thus inducing a homeotropic alignment of the liquid crystal
in the bulk, we have treated the surface of the glass plates by chemically 
adsorbing a surface active layer on them \cite{Cognard81}. 
Prior to the chemical treatment, 
the glass plate is thoroughly cleaned by 
gently washing it with soap and water in order to remove any dirt spots, 
followed by soaking in soapy water for about 60 minutes, then rinsed in 
distilled water and quickly dried, first with clean air, and then baked under
vacuum at 110$^{\circ}$C for about 30 minutes.
We have tested the use of CTAB 
(hexadecyl trimethyl ammonium bromide) and the use of a silane compound
from Merck (ZLI-2510)  to align the liquid crystal molecules. 
In order to treat the plates 
with  CTAB, we dip them into a 0.002\% 
aqueous solution of the surface-active agent. The plates are subsequently
lifted slowly out of the solution in order to avoid precipitation of
CTAB. For the treatment with the silane compound, 
we prepare a saturated solution of
the surface active agent in toluene. The glass plates are dipped in the 
solution for a few seconds and quickly lifted out to avoid creating dry
silane spots. The excess silane is removed from the glass plates by
gently rubbing with lens cleaning tissue. The cells are mounted with the 
desired gap between plates using either nylon thread (for gaps in the
range 100 $\mu$m - 400 $\mu$m) or nickel wire (for gaps smaller than
100 $\mu$m) as spacer. A precision better that 5\%
in the cell gap is obtained this way. The cells are partially
filled
by capillarity, keeping them at a temperature in the nematic range of the
liquid crystal. Since the observations are to be made in the nematic
phase, we make sure that the sample does not go into the isotropic phase
once the
cells are filled. Even though homeotropic alignment of the nematic phase
is achieved faster if the product in the cell is first melted into the 
isotropic
phase, we have observed in some cases that the motion of the N/air
meniscus caused by the melting of the nematic phase crucially alters the 
observable 
phenomena. Sealing of the cell can also be performed to avoid degradation
of the material or to prevent motion of the N/air meniscus.
Different liquid crystal mesophases have been used in our experiments: 
5CB (K15 from Merck),
8CB (K24 from Merck), 
MBBA (from ACROS),
and I52 (from Merck).
Samples with gaps in the range $10 \mu m \leq b \leq 450 \mu m$ have been
used.
\subsection{Preliminary observations}
\label{sec:observations}
%
%
%
%
\begin{figure}[t]
\leavevmode
\vbox{
\epsfxsize = \the\hsize
\epsffile{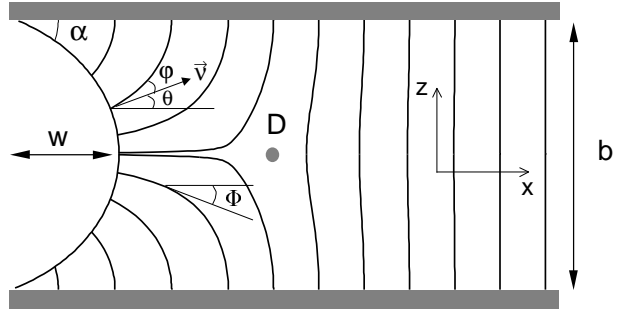}
}
\caption{Cross section of the
translationally invariant distortion caused on the nematic director
field by a boundary with an isotropic medium. The distortion may generate a
$-\pi$-disclination line along the meniscus (D). 
The presence of the isotropic
medium induces the liquid crystal molecules to be anchored
 on the meniscus at
a given angle ($\varphi$) with respect to the local normal ($\vec{\nu}$).
$\theta$ is the angle between the normal and the horizontal axis.
We assume that the meniscus
has a circular profile with radius $R = b/(2\cos(\alpha))$, 
and a wetting angle on the glass
walls $\alpha$. The director field distribution is given, in the
planar approximation, by ${\bf n} = (\cos(\Phi),0,\sin(\Phi))$.
The origin of coordinates is taken to be at the tip of the N/air meniscus.
\label{fig:sample_configuration}
}
\end{figure}
%
%
%
%
\begin{figure}[t]
\leavevmode
\vbox{
\epsfxsize = \the\hsize
\epsffile{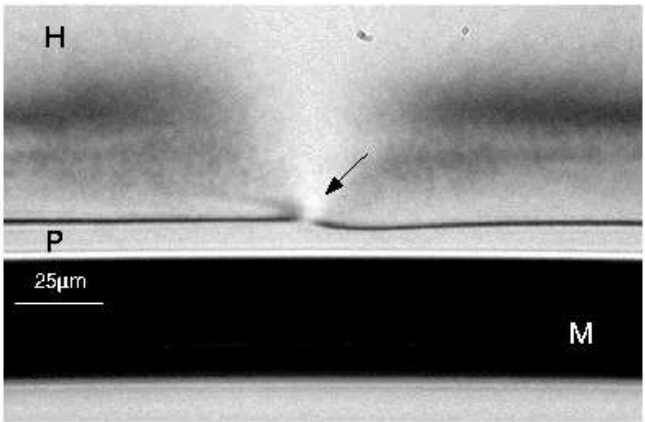}
}
\caption{Disclination line observed from above near a 8CB N/air interface
confined between two parallel glass plates. 
The cell gap is $b = 100 \; \mu$m.
The sample is illuminated with
polarized light. H: homeotropically
aligned nematic phase. P: planar aligned nematic phase between the disclination
line and the meniscus.
 M: N/air meniscus from whose width we
can estimate the wetting angle on the glass plates
(see Fig. \protect{\ref{fig:d_vs_b}}). We can observe a disclination
line that is parallel to the meniscus, except for the region near a 
three-dimensional structure (pointed to by an arrow). 
The presence of such defects suggests that the
director field has a structure that {\sl escapes} the plane formed by the
normal to the meniscus and the perpendicular to the glass plates.
\label{fig:3D-deffect}
}
\end{figure}
When a disclination line is present, 
it appears equidistant from the two glass plate boundaries 
and parallel to the meniscus, suggesting a structure that is translationally
invariant along the interface (see Fig. \ref{fig:sample_configuration}). 
When the sample is observed between crossed polarizers 
(with the polarizers being parallel or perpendicular to the disclination line), 
the region between the disclination line and
the homeotropic bulk reveals the presence of a distortion of the director
field which indicates that it has a component in the direction
of the disclination line (see Fig. \ref{fig:3D-deffect}). The presence
of three-dimensional defects on the disclination line, and the fact that
the position of the line is altered in the vicinity of such defects suggest
that the three-dimensional structure of the director field may 
play a significant
role in the position of the disclination line. As we will see below, however,
a simple planar model for the disclination line (in which the director
field has no component along the line) is able to reproduce 
both qualitatively and quantitatively
our experimental observations on the line position
(see section \ref{sec:model}).
A three-dimensional extension of the
planar model shows that even though the component of the director field 
along the disclination line is not negligible, its effects on the position
of the line are not very significant (see section \ref{sec:model_3D}).

In order to measure
the distance of the disclination line to the meniscus, we always select a
region that is reasonably far from any such defects and that does not appear
distorted by them. For gaps $b < 10\; \mu$m, both the limit of our optical 
resolution and the presence of the three-dimensional defects make our
observations difficult. We are able to observe 
the formation of the disclination line, but we are unable to meaningfully 
measure its distance to the meniscus. Therefore, we will restrict
our measurements to samples with $b > 20\;\mu$m.

\subsection{Equilibrium position of the disclination line}
\label{sec:d_vs_b}
We first studied the 
dependence of the position of the disclination line on the thickness of
the sample. The measurements are performed at
a constant temperature in a region of the nematic phase where temperature
variations of about $1^{\circ}$C
 have a negligible effect on the position of
the disclination line. Typically, this is the situation when the temperature
is more that $1^{\circ}$C above the Nematic/Smectic-A phase transition 
(provided a Smectic-A phase exists) and more that $1^{\circ}$C 
below the Nematic/Isotropic phase transition.%
%
%
%
%
\begin{figure}[t]
\leavevmode
\vbox{
\epsfxsize = \the\hsize
\epsffile{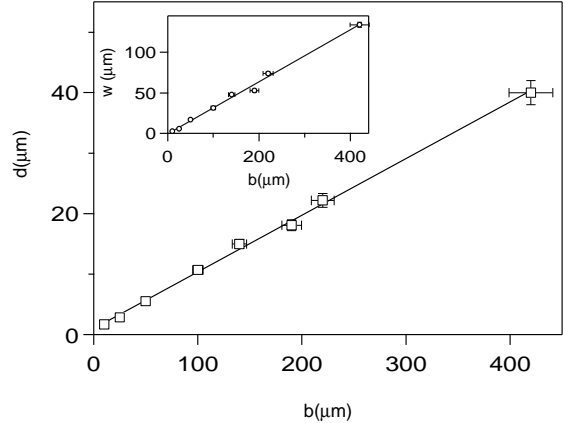}
}
\caption{Distance of the disclination line to the meniscus in a 5CB N/air
interface. The distance is proportional to the gap between the glass plates,
and $d/b = 0.094 \pm 0.005$.
The inset shows how the wetting contact angle of the meniscus at
the glass walls is experimentally measured by assuming a circular profile 
and by measuring the width of the meniscus, 
w (see Fig. \protect\ref{fig:sample_configuration}). 
An average value of w$/b$ is obtained using
measurements with all cell gaps. This average value (slope of the solid line)
is used to compute the contact angle, 
$\alpha = \pi/2 - 2 \tan^{-1} \left( 2\text{w}/b \right )$ 
($\alpha \sim 25^{\circ}$
for this configuration).
\label{fig:d_vs_b}
}
\end{figure}
In Fig. \ref{fig:d_vs_b} we plot the sample-thickness dependence of
the position of the disclination line for 5CB. Our liquid crystal
has $T_{NI} \simeq 34.7^{\circ}$C, and we perform our measurements at
$T \simeq 30.0^{\circ}$C ($T_{NI}-T \simeq 4.7^{\circ}$C).
As we see in Fig. \ref{fig:d_vs_b}, the distance of the disclination line
to the meniscus scales linearly with the cell gap over the range of gaps 
we have explored. 
Clearly, the shape of the meniscus, particularly the wetting angle with
the glass plates, will have a strong influence on the 
position of the disclination line. Assuming a constant radius of curvature
(circular profile), we are able to experimentally estimate the wetting
angle on the plates (see Fig. \ref{fig:d_vs_b}). Given that the N/air
surface energy ($\sim 10^{-3}-10^{-2}$ Jm$^{-2}$)
 is orders of magnitude larger than the nematic anchoring
energy on the free interface
($W \sim 10^{-6}-10^{-5}$ Jm$^{-2}$) \cite{Lavrentovich98}, 
a shape with a circular profile,
which both minimizes the surface area between the two
media and meets the glass plates at a given contact angle, should be 
expected.
Similar results are observed with 8CB, MBBA, and I52,
with a combination of $d/b$ ratios and wetting contact angles $\alpha$ as 
shown in Table \ref{table:lines}.
%
%
%
%
\begin{table}
\caption{Ratio of distance of disclination line to N/air meniscus,
$d$, to the cell gap, $b$, and wetting contact angle on the glass plates,
$\alpha$, for several liquid crystals and surface treatments. The wetting 
properties of the glass plates are noticeably different with the two
different surface treatments we have used.
\label{table:lines}
}
\begin{tabular}{lcccc}
ZLI-2510  & 5CB & 8CB 	& MBBA 	& I52 \\
\tableline
$d/b$  	& 0.094$\pm$0.005 & 0.105$\pm$0.005%
& 0.105$\pm$0.005 & 0.096$\pm$0.005 \\
$\alpha$ & $25^{\circ}$ & $20^{\circ}$ & $25^{\circ}$ & $8^{\circ}$\\
\end{tabular}

\begin{tabular}{lcccc}CTAB & 8CB & MBBA\\
\tableline
$d/b$  	& 0.100$\pm$0.005 & 0.100$\pm$0.005\\
$\alpha$ & $10^{\circ}$ & $10^{\circ}$\\
\end{tabular}

\end{table}
It is customary to construct the dimensionless parameter $Wb/K$, which
compares the length scale imposed by the confined geometry
(in our case the cell gap, $b$), with
an intrinsic length scale of the material, $K/W$.
Note that throughout this article we will suppose that the anchoring
energy on the glass plates (which induces the homeotropic orientation
of the bulk) is large compared to $W$, so that the presence of the
free interface does not alter the anchoring condition on the plates. 

The perturbation caused on the homeotropic director field configuration
by the presence of the free interface will be less apparent for
samples thinner than $K/W$.
One expects the
formation of a disclination line for $Wb/K >> 1$, when surface anchoring
effects dominate and impose a rigid anchoring angle. 
If $Wb/K << 1$, however, elastic effects will 
dominate and a balance between elastic and surface anchoring torques, 
rather that a rigid anchoring angle, is
expected to determine the anchoring condition at the meniscus. 
In this case, the distortion
caused by the free interface will be less important, and the 
equilibrium structure may be one without a disclination line. 
The results reported in Fig. \ref{fig:d_vs_b} and Table \ref{table:lines},
namely, a disclination line whose position is completely determined by the
geometry of the system, prove that the only relevant length scale 
is $b$, the cell gap. We are clearly in a regime with $Wb/K >> 1$,
or {\sl infinite} anchoring energy, where the anchoring angle on the meniscus
is rigidly set. If we were able to make clear observations with a small
enough sample thickness (typically $b<10\; \mu$m), 
we should presumably see the end of the
$d \propto b$ regime, and observe the collapse of the disclination line.

In the next section, we outline a model to relate the equilibrium
position of the disclination line to the different experimental
parameters. Our model predicts the disclination line to start 
collapsing noticeably for
$Wb/K < 5$ (see Fig. \ref{fig:numeric_basic}), which we may use to give
a lower bound for the anchoring energy, $W$. 
We have confidently measured the position of the disclination line 
for  $b > 25\; \mu$m, and still observe the same  $d/b$ ratio (within
error bars). We can, thus, estimate
$W > 5 K/b = 1.1\; 10^{-6}\;\text{J}\text{m}^{-2}$ 
($K_{3} = 5.5\; 10^{-12}\;\text{N}$ for 5CB at $T_{NI}-T = 5.0^{\circ}$C
\cite{Madhusudana82}).
A similar estimate is obtained from our measurements with 8CB.

%
%
%
%
\begin{figure}[t]
\leavevmode
\vbox{
\epsfxsize = \the\hsize
\epsffile{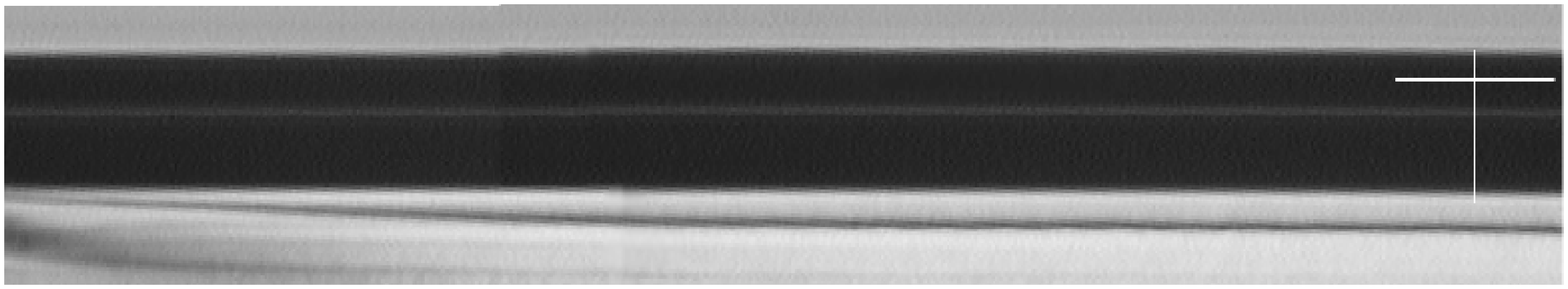}
}
\vbox{
\epsfxsize = \the\hsize
\epsffile{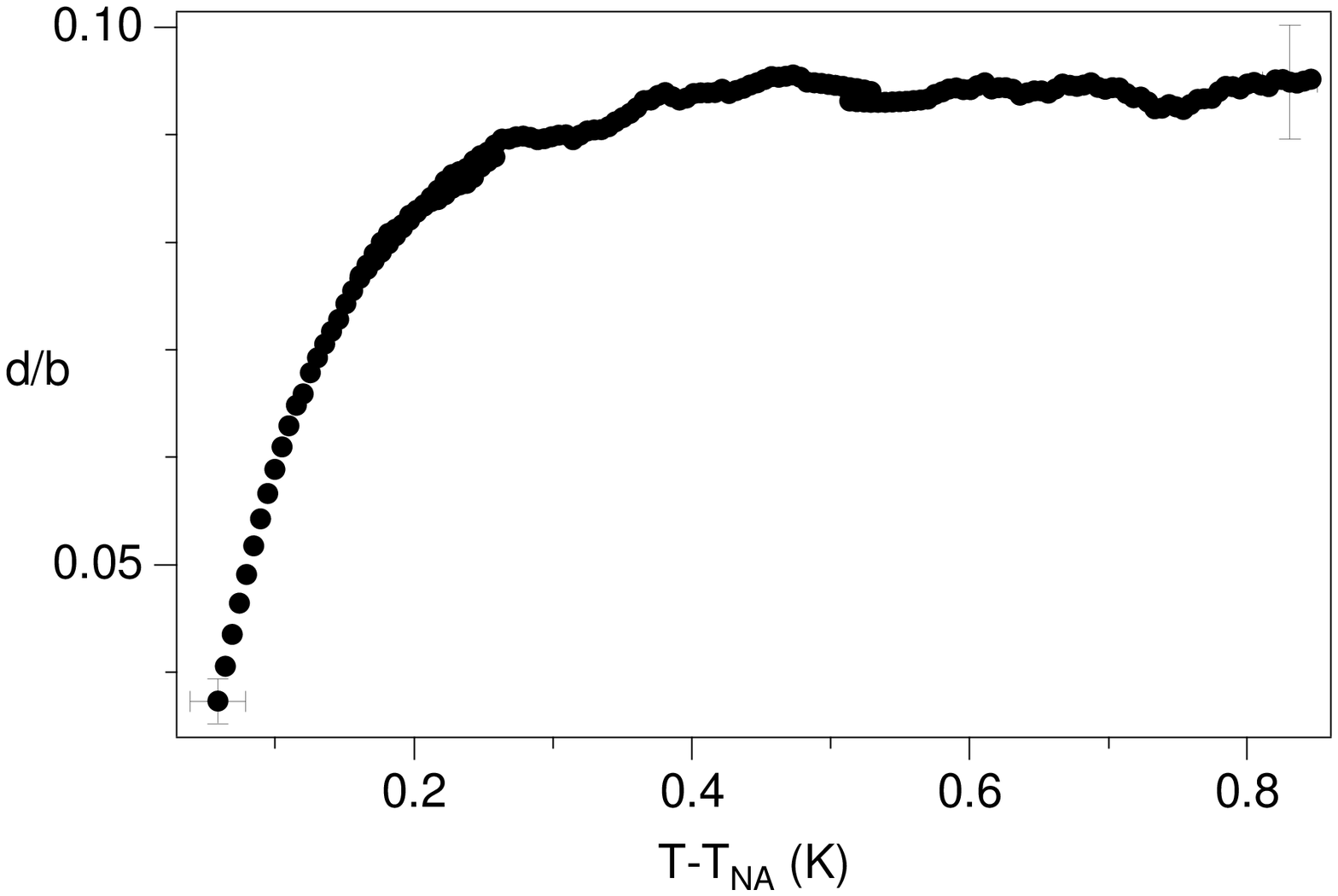}
}
\caption{Collapse of the disclination line close to $T_{NA}$ 
for 8CB in the presence of a N/air interface
with a sample of thickness $b = 25\;\mu$m.
The imposed temperature gradient is $38.6^{\circ}\text{C}\,\text{cm}^{-1}$.
Top: picture of the sample from above, observed between partially crossed
polarizers. The length of the vertical line segment is $10\;\mu$m and that
of the horizontal segment is $15\;\mu$m. $T_{NA}$ occurs at the left
edge of the picture.
Bottom: the distance of the disclination line to the meniscus is
measured as a function of the distance to the transition point, and
converted to $T-T_{NA}$ by means of the known temperature gradient.
The disclination line reaches its
plateau position at about $0.5$ K above $T_{NA}$.
\label{fig:nearTNA}
}
\end{figure}
In an attempt to observe experimentally the collapse of the disclination line
due to small values of $Wb/K$, we have considered altering the value of $W$
by replacing the air interface with an interface
with another isotropic medium, such as an isotropic fluid or glass wires
with different surface treatments.
Even though we have observed an effect of the different interfaces on
the anchoring conditions, the configurations had poor reproducibility
and dubious stability. We also tried to increase $Wb/K$ by approaching 
the nematic/Smectic-A phase transition temperature. Indeed, we know from
previous 
work that the bend elastic constant $K_3$ diverges at this transition 
\cite{Madhusudana82,Davidov79,Garland94} 
so that the disclination line energy also diverges.  
The experiment was 
performed by placing a $25 \; \mu$m-thick sample of 8CB inside a constant
temperature gradient parallel to the liquid-crystal/air interface 
(see Fig. \ref{fig:nearTNA}). The immediate observation is that  while in
the N phase
a disclination line is observed, it disappears as we enter the SmA phase.
More precisely, 
$d/b$ shifts from its asymptotic value (close to $0.1$) 
when $T-T_{NA}<0.3^{\circ}$C and typically equals  
$0.05$ at $T-T_{NA} \simeq 0.1^{\circ}$C. At this temperature $K_3 \simeq 
4\; 10^{-11}\;\text{N}$ while $d/b=0.05$ when $Wb/K \simeq 1$ (see the
following section). 
This allows us to roughly estimate W in 8CB: $W \simeq 1.5\;
10^{-6}\;\text{J}\text{m}^{-2}$.
\section{Model}
\label{sec:model}
In order to better relate the phenomena described in the previous section
to the experimental parameters,
we modeled the director-field configuration in a
confined nematic geometry in the presence of a free interface. 
Even though our observations show
a three-dimensional director-field structure around the disclination line
(see section \ref{sec:observations}), we propose a simple,
two-dimensional model with a planar disclination line
(the director field is assumed to be perpendicular
to the line). This model has allowed us to obtain numerical estimates
for the position of the disclination line that
match very well the experimental observations. Finally, we 
consider a more accurate model, where the three-dimensional structure
of the director field
is taken into account. We verify that, in fact, the
predictions from the planar model quite accurately predict the position
of the disclination line.
\subsection{Equations}
\label{sec:basic_equations}
If we consider that the distortion caused by the presence of the free
interface is translationally invariant along the
interface, and ignore the component of the director field along
the disclination line (planar disclination line),
then a typical director field configuration will appear as shown
in Fig. \ref{fig:sample_configuration}, and it will be characterized
by the angular field $\Phi(x, z)$, with ${\bf n} = (\cos(\Phi),0,\sin(\Phi))$.
Given the two-dimensional nature of this approximation, there will be
no contribution from the twist of the director field to the elastic
free energy (${\bf n}\cdot{\bf curl}\;{\bf n} =0$).
Likewise, the saddle-splay term ($K_{24}$) will vanish identically
\cite{Frank58,Neyring71}.
Moreover, for the experimental conditions described in the previous 
section in which the samples are kept at a temperature away from both the N/I
and N/SmA transitions, the splay and bend elastic constants take on roughly
the same values, $K_{1} \simeq K_{3}$. A final approximation is
made in which the mixed splay-bend term ($K_{13}$) is neglected
\cite{Neyring71,Yokoyama97}. 
With all this, we can write,  
\begin{equation}
\label{eq:frank_energy}
f = \frac{1}{2} K \left[ ({\bf div}\;{\bf n})^{2} + 
({\bf n}\times{\bf curl}\;{\bf n})^{2} \right],
\end{equation}
which gives
\begin{equation}
\label{eq:frank_energy_2d}
f = \frac{1}{2} K \left[ \left(\frac{\partial \Phi}{\partial x} \right)^{2}
                          +\left(\frac{\partial \Phi}{\partial z} \right)^{2}
                    \right],
\end{equation}
where $K_1 \simeq K_3 = K$.
The glass walls are taken
to be at $z = \pm b/2$, and we assume that a strong anchoring fixes
$\Phi = \pm\pi/2$ there.

The interaction with the N/air meniscus contributes a surface energy
that is customarily approximated by
\begin{equation}
\label{eq:surface_energy}
\gamma = \gamma_0 - \frac{W}{2} \cos^{2}(\varphi - \varphi_0),
\end{equation}
with $\varphi = \Phi - \theta$. Here, 
$\theta$ is the angle between the local normal to the meniscus 
and the horizontal direction, and $\varphi_0$
introduces a favored anchoring angle between the director field and the
local normal to the interface for which the surface energy density is
minimum (see Fig. \ref{fig:sample_configuration}). 

The energy introduced in the system by the presence of the N/air
interface will be
\begin{equation}
E = \frac{1}{2} K \int ({\bf grad}\Phi)^2 \;dv
- \frac{1}{2} W \int \cos^2(\varphi-\varphi_0)\; dS,
\end{equation}
where the volume integral spans over the bulk nematic phase and the
surface integral spans over the N/air meniscus.

We will seek the director-field distribution (characterized by $\Phi$) by
minimizing the energy functional with respect to $\Phi$. Integrating
by parts and imposing that the director field be homeotropic both
on the glass boundaries and far from the interface, one obtains
\begin{mathletters}
\label{eq:configuration}
\begin{eqnarray}
   \Delta \Phi \; = \; 0 &,& \text{ in the bulk}, \\
   \label{eq:anchoring_bc}
   K\frac{\partial \Phi}{\partial \vec{\nu}} \; = \; 
     \frac{W}{2} \sin[2(\varphi-\varphi_0)] &,& \text{ on the free interface.}
\end{eqnarray}
\end{mathletters}
 
Eq. (\ref{eq:anchoring_bc}) establishes a torque balance at the 
interface (here, $\partial/\partial \vec{\nu}$ 
represents the derivative along the local normal direction, and
$\theta$ is the angle between the local normal to the interface ($\vec{\nu}$)
and the horizontal direction). The shape of
the interface is completely determined by the experimentally measured
wetting contact angle (see Fig. \ref{fig:sample_configuration}).
These equations can be made dimensionless by rescaling all length scales by $b$
and defining the anchoring parameter $Wb/K$, which contains all the material
properties. With this, one would expect the distance from the disclination
line to the meniscus to have a functional dependence of the form
$d = b f(Wb/K)$. In the strong anchoring regime ($Wb/K \rightarrow \infty$),
$\Phi = \theta + \varphi_0$ will be rigidly set on the interface, 
and $d \propto b$ is expected.
Note that there strictly exists in our model a discontinuity in the director 
field at the tip of the meniscus when $\varphi_0 > 0$ is rigidly imposed at 
the interface. This may lead to values of $d$ smaller than the real ones.
The use of eq. \ref{eq:anchoring_bc} with a large, but finite, value of $Wb/K$
effectively removes this discontinuity
and shows that the error introduced by the {\sl infinite} anchoring 
approximation is very small.
%
%
%
%
\begin{figure}[t]
\leavevmode
\vbox{
\epsfxsize = \the\hsize
\epsffile{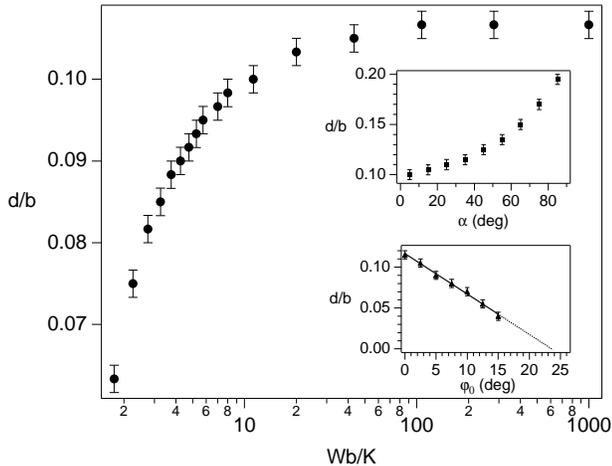}
}
\caption{
Expected position of the planar disclination line, $d$, in terms of
the different experimental parameters and expressed as the ratio
$d/b$.
($\bullet$): fixing $\alpha = 25^{\circ}$ and $\varphi_0 = 0^{\circ}$,
the position of the disclination line as a function of the anchoring
strength parameter, $Wb/K$, is calculated. $d/b$
reaches a limiting value as $Wb/K$ is increased (limit of
{\sl infinite} anchoring). In the insets we show the ratio 
$d/b$ (in the  $Wb/K \rightarrow \infty$ limit) as a function of
the wetting angle, $\alpha$ ($\Box$), in the $\varphi_0 = 0$ configuration,
and as a function of the anchoring angle, $\varphi_0$ ($\triangle$),
for a wetting angle $\alpha=25^{\circ}$. Note that our model predicts that the 
disclination line will approach the meniscus roughly linearly with
increasing $\varphi_0$, and that one can extrapolate the value of $\varphi_0$
at which the line would collapse (dashed line).
\label{fig:numeric_basic}
}
\end{figure}
\subsection{Numerical model}
In order to find the configuration which  minimizes the total energy,
we proceed to solve eq. \ref{eq:configuration} by relaxing the fixed
boundary problem obtained by setting the position of the disclination
line at a distance $d$ from the meniscus. Therefore, 
$\Phi(x,0^{\pm}) = \pm \pi/2$ is set for $x > d$, while $\Phi(x,0)=0$ is
set for  $0 < x < d$.
In order to avoid the 
$\pi$-discontinuity at $z = 0$, we restrict our calculation to the
$z>0$ region. The value of the wetting angle (which can be determined
experimentally, see Fig. \ref{fig:d_vs_b} and Table \ref{table:lines})
and the value of the favored anchoring angle, $\varphi_0$, together
with the condition of homeotropic configuration both far from the meniscus
and on the glass walls,
completely determine the director field distribution. Typically,
we use discretization grids of about 400x200 points. 
Once the configuration is obtained, we compute its total 
energy and then minimize this energy with respect to the 
configurations obtained by changing the value of $d$. Note that we
neglect the contribution of the core of the disclination to the
total energy. We assume that it will be constant, independent of
$d$ and, therefore, will not be relevant in the determination of the
configuration with the least energy.
In Fig. \ref{fig:numeric_basic},
we show the results of these calculations. First, we explore
$d/b$ as a function of $Wb/K$. This corresponds to the measurements
performed in section (\ref{sec:d_vs_b}). We observe that the model 
predicts that
$d/b$ converges asymptotically to $d/b = 0.105 \pm 0.003$ as  
$Wb/K \rightarrow \infty$, for $\alpha = 25^{\circ}$, and  $\varphi_0 = 0$. 
Note that this is, within error bars, consistent with the results 
reported in Table \ref{table:lines}. Only our measurements with 5CB
(shown in Fig. \ref{fig:d_vs_b}) are slightly smaller that the prediction
from our model.
Our numerical calculations predict as well that, 
as $Wb/K$ gets smaller, $d/b$ starts to
decrease. In order for $d/b$ to decrease more than, say, 10\%
below the infinite anchoring position (which would be experimentally
measurable with our optical resolution) one would need $Wb/K < 5$
(see Fig. \ref{fig:numeric_basic}).
We have also explored the effect that $\alpha$ and  $\varphi_0$ have
on the disclination line. We can see in  Fig. \ref{fig:numeric_basic}
that $d/b$ increases with increasing values of $\alpha$.
Indeed, if we held $d/b$ fixed, a flatter meniscus
(larger $\alpha$) would require larger distortions of the director
field in order to match both the strong anchoring on the glass walls
and on the meniscus. This would increase the elastic energy, which
can be prevented by increasing $d$. The experimental observations 
reported in Table \ref{table:lines} are consistent with this result,
even though the dispersion in the data, combined with  
the fact that in the $\alpha <25^{\circ}$ region $d/b$ is
not very sensitive to changes in $\alpha$, do not allow for any
conclusive trend to be extracted from our measurements. 
On the other hand, we may exploit
the modeled effects of the anchoring angle, $\varphi_0$,
to study the known phenomenon of temperature-induced anchoring
transitions (see section \ref{sec:anchoring_transition}). 
We observe (see  Fig. \ref{fig:numeric_basic}) that, within our
numerical resolution, $d$ decreases roughly linearly with increasing
values of $\varphi_0$. Using this result, one can immediately
correlate changes in the position of the disclination line with 
changes in the anchoring condition, provided $W$ remains unchanged.
Observe that, extrapolating the calculations reported in
Fig. \ref{fig:numeric_basic}, we can expect that, even in the presence of a 
strong anchoring,
a configuration without a disclination line may be observed, provided
$\varphi_0$ is large enough ($\varphi_0 > 25^{\circ}$ for the configuration
in Fig. \ref{fig:numeric_basic}). 
This is the reason why 
a real disclination line is not formed spontaneously near a nematic/isotropic
front, since on that interface large values for $\varphi_0$ are reported
($\varphi_0 \simeq 48^{\circ}$ for 8CB \cite{Faetti84}). We recall that the
anchoring length at this interface $K/W$ is too small (of the order of $5\mu$m
with $W \sim 8.5\; 10^{-7}$ Jm$^{-2}$ \cite{Faetti84} and $K_3 \simeq 
4\; 10^{-12}\;\text{N}$ \cite{Madhusudana82})to explain alone 
the absence of a real disclination line in thick samples ($d\sim
50-400\mu$m). We 
also emphasize that the formation of planar triangles in directional
solidification 
\cite{Oswald87}
requires that a dust particle crosses the N-I interface, suggesting the
disclination
line is not real but virtual (its core lies in the isotropic liquid). This
observation 
fully agrees with our predictions.

\subsection{Three dimensional effects}
\label{sec:model_3D}
%
%
%
%
\begin{figure}[t]
\leavevmode
\vbox{
\epsfxsize = \the\hsize
\epsffile{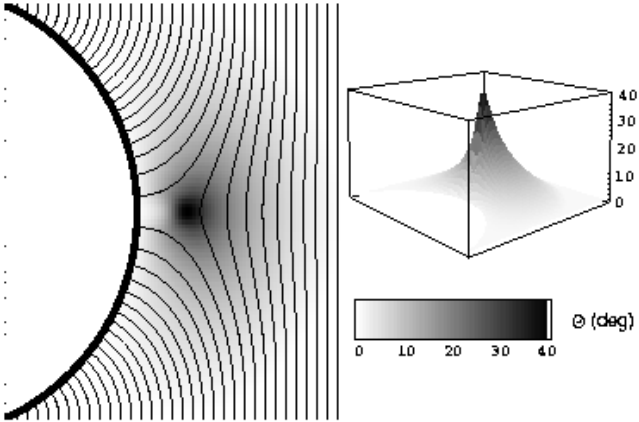}
}
\caption{
Three dimensional structure of the director field near the N/air meniscus.
The field lines are tangent to the planar component of the director field,
and the shading represents the angle between the director field
and the plane, $\Theta$, where $\Theta = 0$ is the planar case.
Darker areas correspond to larger third-dimensional component.
The same angle is plotted as a surface plot inside the box. 
The escaped component diverges in the core of the disclination
line. The core of the disclination is centered at a distance 
$d \simeq 0.11 b$, in agreement with the planar model and with our
observations.
\label{fig:3d-model}
}
\end{figure}
As we have seen in section (\ref{sec:observations}), 
our experimental observations 
reveal a three-dimensional structure in the director field in the vicinity 
of the disclination line. The model described above 
assumes a planar configuration of the
director field, in which the component along the disclination line is
ignored. In this section, we study the modifications introduced by a
three-dimensional model of the director field. 

We have used the full Landau-Ginzburg-de Gennes tensorial expansion 
for the elastic free energy density
in power series of the quadrupolar order parameter
(which we define as $Q_{ij} = (Q/2)(3n_i n_j -\delta_{ij})$, assuming a
uniaxial system).
We take $K_1 = K_3 = 2 K_2$, which is
an excellent approximation for 5CB and 8CB in the region of the
nematic phase we study. 
Moreover, the saddle-splay
term is included as well ($K_{24}=-K_1 $) and the mixed splay-bend term is
set to zero ($K_{13}=0$)\cite{Yokoyama97}. 
We have only considered the case of rigid 
homeotropic anchoring,
in which the liquid crystal molecules are forced to anchor on the meniscus
at $\varphi_0 = 0^{\circ}$. A detailed description of the model and the
numerical methods involved are given in \cite{Baudry98}.

Indeed, we obtain an equilibrium configuration where the director field
has a significant projection in the direction of the disclination line,
especially in the vicinity of the line (see Fig. \ref{fig:3d-model}). 
Surprisingly, though, when the
position of the disclination line is estimated, the result is, within
error bars, the same as the one found with the much simpler planar model.
Therefore, the three-dimensional calculations validate the use of the
planar model to predict the position of the disclination line as a function
of the relevant physical parameters.

\section{Effect of the anchoring angle}
\label{sec:anchoring_transition}
We have seen on
section \ref{sec:model} that, based on the results from our computations,
we expect the disclination line to approach the meniscus as the
anchoring angle is larger than zero and that, presumably, configurations
without a disclination line might be possible, provided the anchoring
angle is large enough.
It is well-known that the equilibrium anchoring angle of the nematic molecules
at a N/air interface can depend on temperature 
\cite{DeGennes,Bouchiat71,Sonin95}.
We have performed
measurements with MBBA and I52, both of which are known to have such a 
behavior, in order to study the effect the anchoring angle has on the formation
of a disclination line.
Samples are prepared as explained above and placed in a temperature-regulated
oven and the position of the disclination line as a function of
temperature is measured.

\subsection{Continuous anchoring angle transition in MBBA and I52}
\begin{figure}[p]
\leavevmode
\vbox{
\epsfxsize = \the\hsize
\epsffile{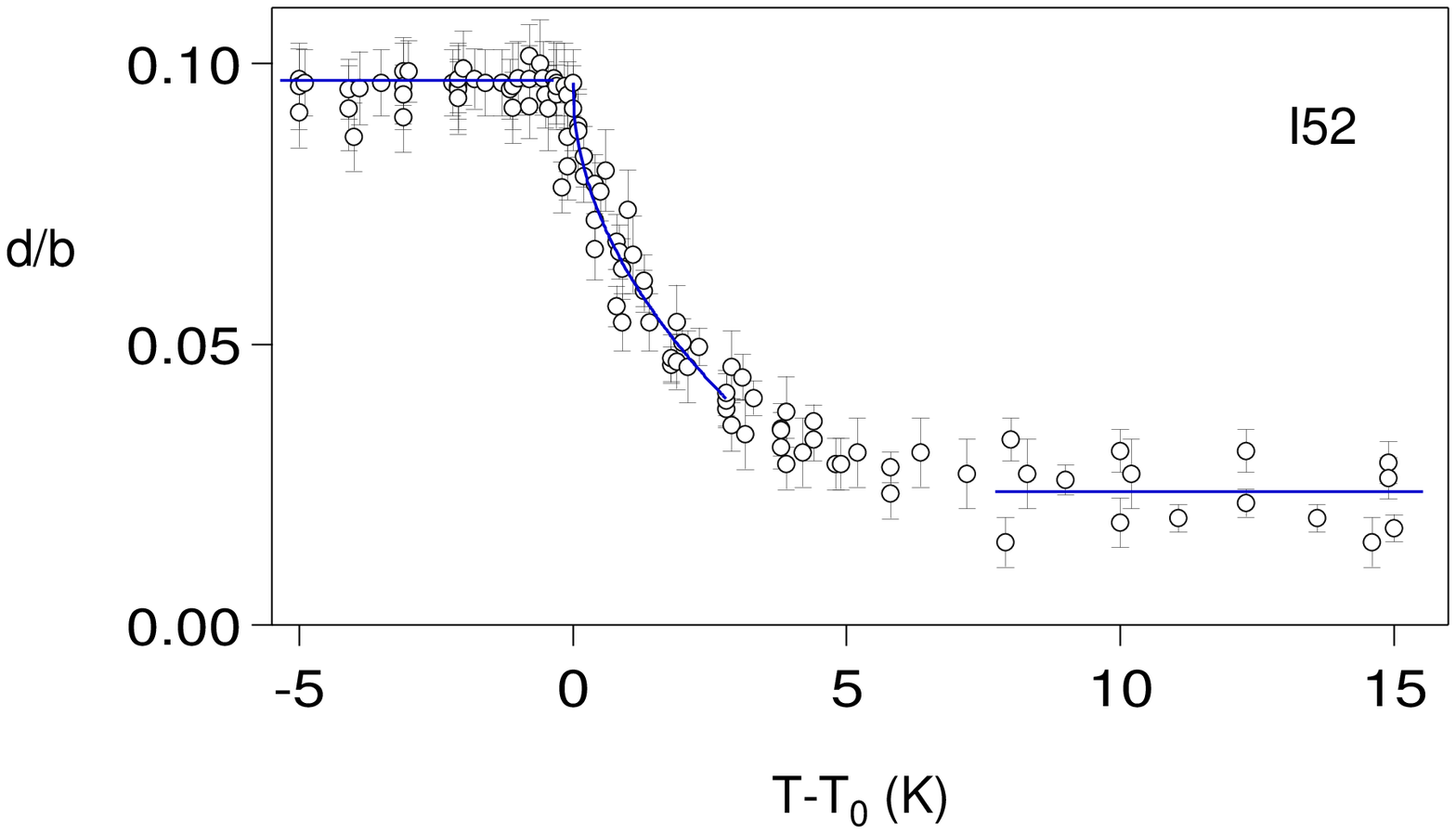}
}
\vbox{
\epsfxsize = \the\hsize
\epsffile{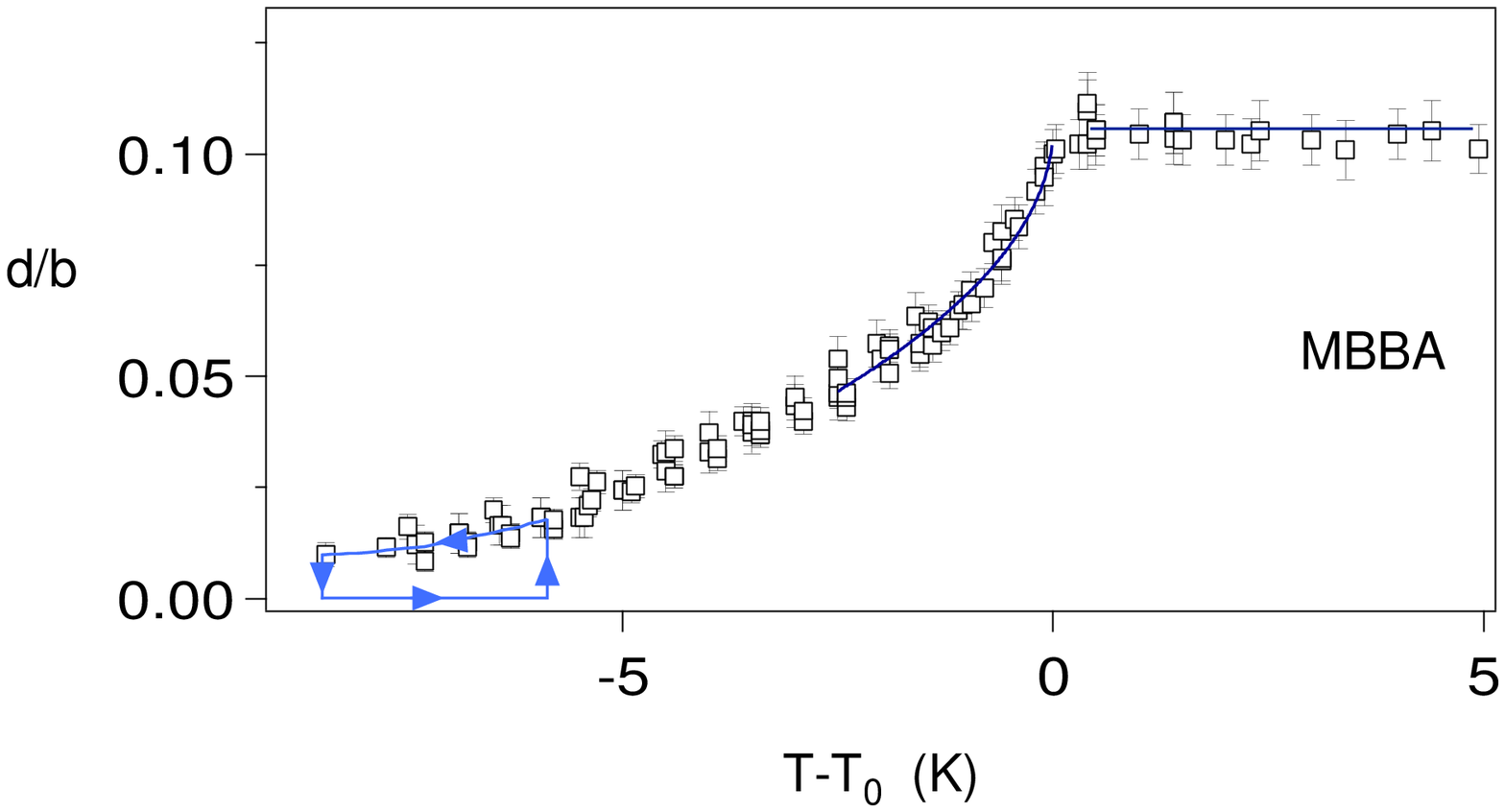}
}
\caption{
Distance of the disclination line to the N/air meniscus ($d$), 
rescaled by the sample thickness ($b$) as a function of the
departure from $T_{0}$, the temperature above or below which
the disclination line starts approaching the meniscus, 
showing a continuous transition between two different configurations
presumably due to an anchoring angle transition.
Top: measurements performed with I52, with sample thicknesses $b=50\;\mu$m,
$110\;\mu$m, $140\;\mu$m, and $220\;\mu$m. 
Bottom: measurements performed with MBBA, with sample thicknesses
$b=110\;\mu$m, $190\;\mu$m, and $220\;\mu$m.
The departure from the  $\varphi_0 = 0$  configuration 
($d/b \simeq 0.1$) is compatible with $d/b \propto |T-T_0|^{1/2}$ 
(solid curve). 
The solid horizontal lines are guides for the eye in the plateau 
regions, where $d/b$ does not change with $T$.
In both systems, heating and cooling cycles are indistinguishable,
as long as the disclination line does not collapse (MBBA case).
The diagram in the low temperature region illustrates
the hysteretical collapse of the disclination line observed for this
product.
\label{fig:d_over_b_vs_T}
}
\end{figure}

For I52, the disclination line starts to approach continously the 
N/air meniscus as we {\sl increase} the temperature beyond 
$T_0 \simeq 25 ^{\circ}$C, until it reaches a final configuration where the 
distance remains unchanged by further increasing the temperature. No hysteresis
is observed as we alternate heating and cooling cycles, and the transition 
temperature is independent of sample thickness. In fact, measurements
performed with different values of the cell gap are collapsed by rescaling
the distance of the disclination line with the cell gap 
(see Fig. \ref{fig:d_over_b_vs_T}). This suggests that what we are
seeing is not due to changes in the anchoring energy, since its influence
in the position of the disclination line is introduced by the parameter
$Wb/K$, which depends on the sample thickness.
%
%
%
%
For MBBA, the distance from the disclination line to the N/air meniscus
starts to decrease continuously as we {\sl lower} 
the temperature below $T_0 \simeq 26 ^{\circ}$C. Just as with I52,
no hysteresis is observed here.
The transition temperature is slightly different for different
samples, however, and it may depend on sample thickness, 
even though we have been
unable to find any meaningful trend. Shifting the temperature axis with
the value of $T_{0}$ for each experiment, and rescaling the
position of the disclination line by $b$ effectively 
brings onto a single curve the data 
taken with different sample thicknesses (see Fig. \ref{fig:d_over_b_vs_T})
suggesting as well that we are always in the {\sl infinite} anchoring
regime.

Both for I52 and for MBBA,
we see that the departure from the upper plateau is well-described
by a continuous (or supercritical) bifurcation of the form 
$d/b \propto |T-T_{0}|^{1/2}$, which is analogous
to what has been observed elsewhere for anchoring-angle transitions
\cite{Sonin95}. Fitting such an analytical profile to our data, we can get
a more precise estimate for the transition temperature, $T_{0}$.

\begin{figure}[t]
\leavevmode
\vbox{
\epsfxsize = \the\hsize
\epsffile{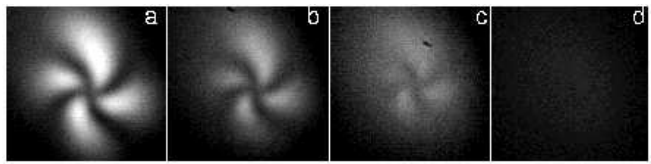}
}
\caption{A nematic microdroplet of I52 is placed on a substrate that induces a
homeotropic anchoring and it is observed between crossed polarizers,
at different temperatures: $T=34.5^{\circ}$C (a), $T=28.2^{\circ}$C (b),
$T=26.2^{\circ}$C (c), and  $T=25.2^{\circ}$C (d). The temperature of
(a) is well in the lower plateau on Fig. \protect\ref{fig:d_over_b_vs_T}.
The observation of a point defect on the free surface suggests a tilted
anchoring of the molecules. As the temperature is reduced, the distortion
decreases until we lower $T$ below $T_0\simeq 25.5^{\circ}$C so that
we are in the upper plateau on Fig. \protect\ref{fig:d_over_b_vs_T}.
The distortion has vanished, suggesting we have a homeotropic anchoring at the
free surface.
\label{fig:I52_drop}
}
\end{figure}

Note that, even though a second stable position for the disclination line
is observed for I52, the line keeps approaching the meniscus as the
MBBA samples are cooled to lower temperatures, until the disclination line
finally collapses. As we describe below, there is hysteresis in this 
collapse, and the dynamics that leads to it is reminiscent of a first-order
transition.
Based on the observations reported above and on the results found in the 
literature, we would like to argue that the changes we see in the position
of the disclination line when we vary the temperature of the samples
can be accounted for by changes in the anchoring angle
at the interface. The existence of an anchoring angle transition is 
easily demonstrated by observing between crossed polarizers
a micro-droplet of the liquid crystal placed on a substrate that induces
a homeotropic anchoring (see Fig. \ref{fig:I52_drop}). 
For temperatures at which the disclination line
is at the upper plateau in 
Fig. \ref{fig:d_over_b_vs_T}, 
we have observed no point defects in the droplets. 
As we vary the temperature entering the region where a transition in the
position of the disclination is observed in Fig. \ref{fig:d_over_b_vs_T},
a characteristic point defect structure starts to become manifest, and
the distortion in the director field increases in the temperature region
where the disclination line is closest to the meniscus.
%
%
%
%
%
%
%
%
\begin{figure}[t]
\leavevmode
\vbox{
\epsfxsize = \the\hsize
\epsffile{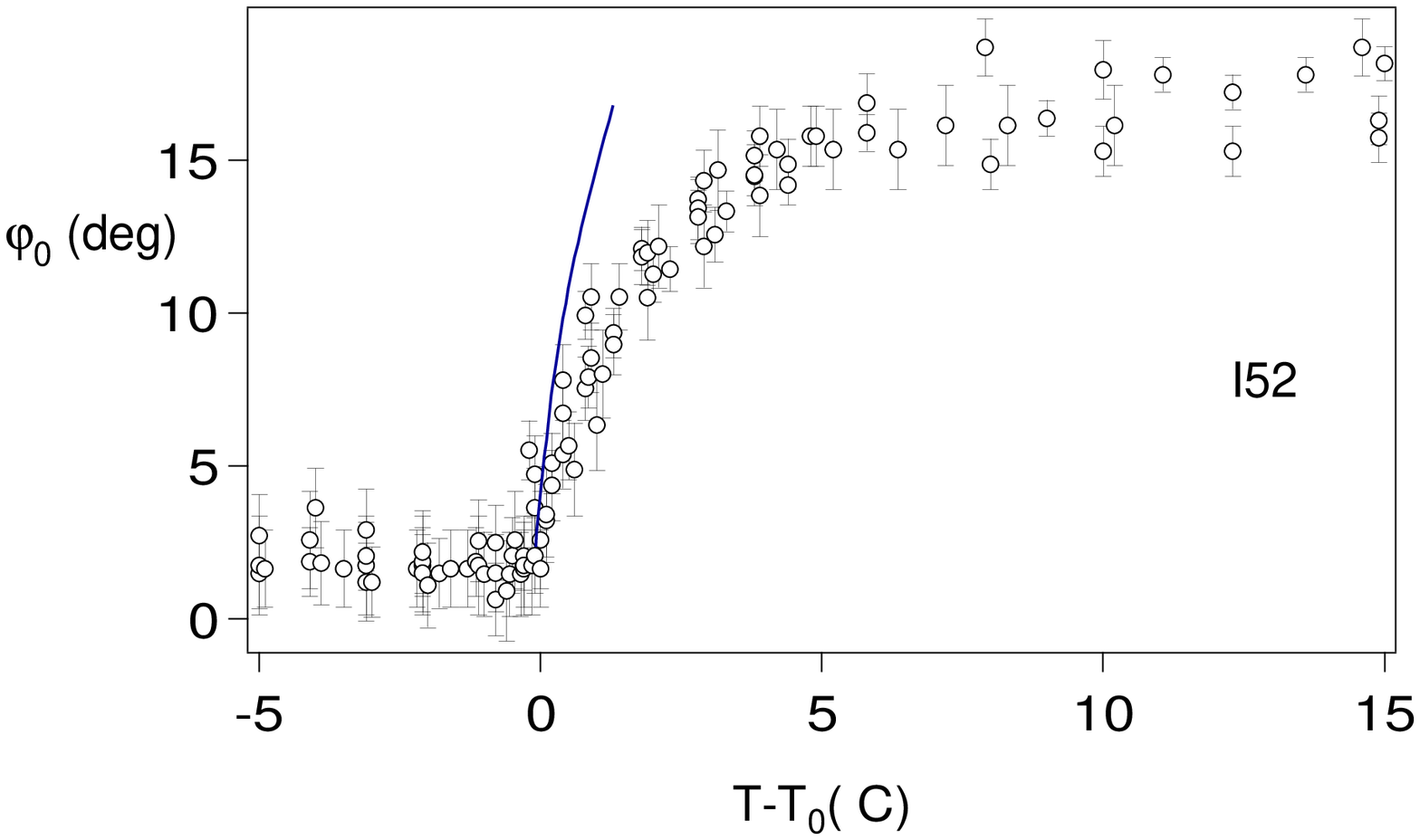}
}
\vbox{
\epsfxsize = \the\hsize
\epsffile{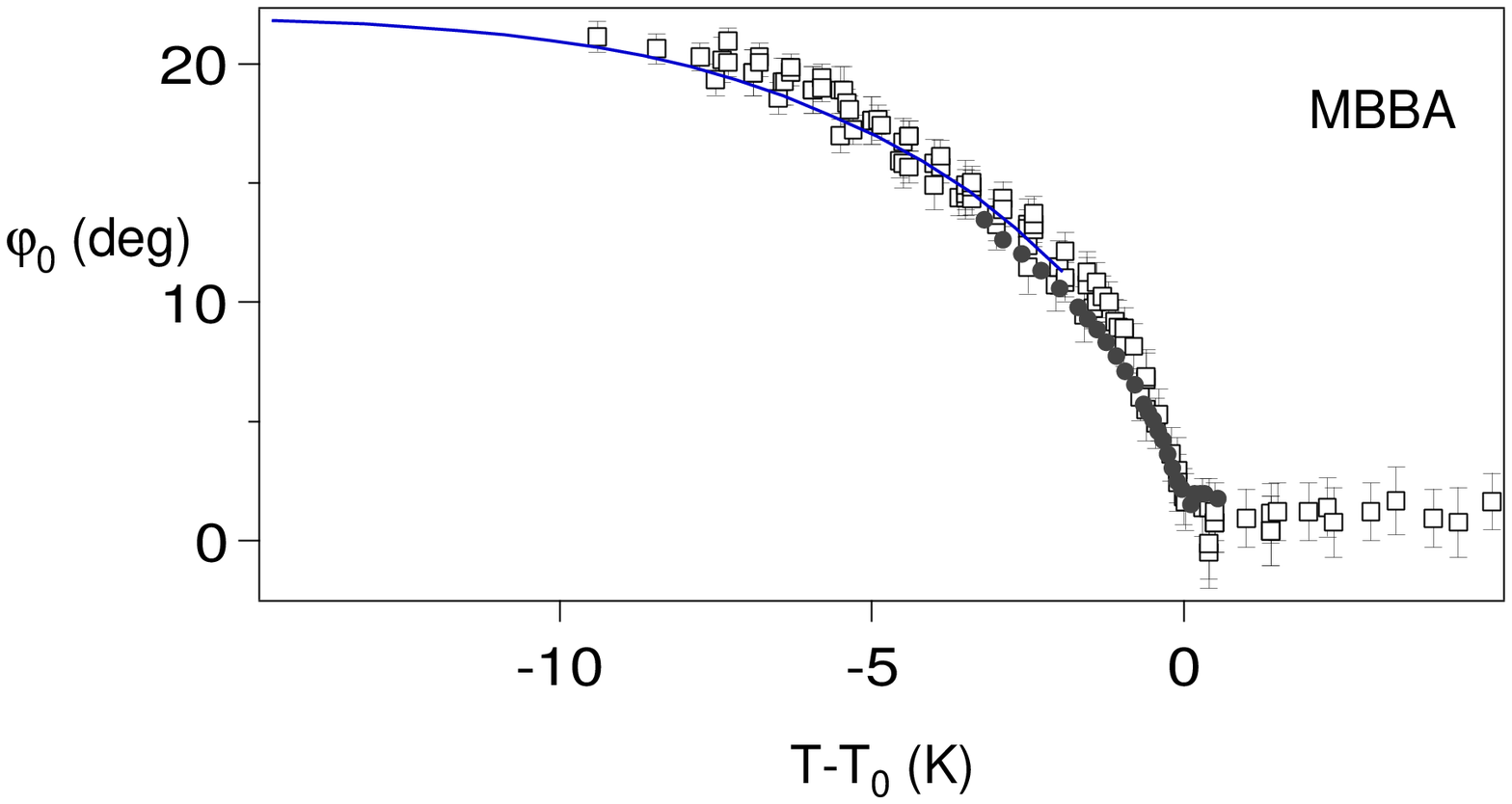}
}
\caption{
Temperature-induced orientational transition of the anchoring angle at
the N/air interface where the angle has been estimated by
transforming the $d/b$ vs. $T-T_{0}$ data on 
Fig. \protect\ref{fig:d_over_b_vs_T} with our numerical model.
Top: results corresponding to the data obtained with I52 ($T_0\simeq
25^{\circ}$C)
and measurements reported by Sonin {\sl et al.} 
\protect\cite{Sonin95} (solid line).
Bottom: results corresponding to the data obtained with MBBA ($\Box$) 
($T_0\simeq 26^{\circ}$C),
results reported by Sonin {\sl et al.} ($\bullet$), and results
reported by Bouchiat {\sl et al.} 
\protect\cite{Bouchiat71} (solid line). 
\label{fig:phi_vs_T}
}
\end{figure}
We have used our planar model for the
structure of the director field near the N/air interface in order
to relate the position of the disclination line to the anchoring angle
(see Fig. \ref{fig:numeric_basic}).
The numerical model considers a planar disclination line and a 
strong anchoring condition at the meniscus with an anchoring angle
that can be different from the homeotropic orientation 
($\varphi_0=0$). Note that our numerical model assumes $K_1 \simeq K_3$.
For MBBA, in the relevant temperature range ($T\simeq 25^{\circ}$C), 
one finds in the literature that $0.84<K_3/K_1<1.4$. The same information,
however, is not available for I52. The fact that our planar model
successfully predicts the position of the disclination line in the homeotropic
anchoring situation is a good indication that the above approximation is
valid. 
Moreover, in the vicinity of the continuous anchoring angle transition,
the angular dependence of the surface energy may have its leading order
term cancelled. Indeed, the surface energy
in the vicinity of the second order anchoring angle transition can be
expanded as
\begin{equation}
\label{eq:near_anchoring_transition}
\gamma(\varphi) = \gamma_0 + \frac{W_0}{2}(T-T_0)\varphi^2
+ B \varphi^{4} + \dots,
\end{equation}
where $W_0$ and $B$ are two positive constants.
The equilibrium angle is $\varphi_0 = 0$ for $T \geq T_0$ and 
$\varphi_0 =\sqrt{W_0(T-T_0)/B}$ for $T<T_0$ (MBBA case).
The torque originated by an anchoring at an angle $\varphi > 0$
at $T = T_0$ is 
\begin{equation}
\frac{\partial\gamma}{\partial\varphi} = W_0 (T-T_0)\varphi
+ 4 B \varphi^{3} + \dots.
\end{equation}
We see that, at $T=T_0$, the leading order term vanishes. We will assume
that the first non-vanishing term, $4 B \varphi^{3}$ still increases
steeply when $\varphi >0$, 
so that the anchoring angle remains close to its value at equilibrium.

The numerical calculations show that the dependence of the
ratio $d/b$ on the anchoring angle can be roughly approximated by a 
straight line,
whose slope depends on the wetting angle of the meniscus on the glass
plates (see Fig. \ref{fig:numeric_basic}).  
Using our numerical results, we have been able to
transform the $d/b$ vs. $T-T_{0}$ data in Fig. \ref{fig:d_over_b_vs_T}
into $\varphi_0$ vs. $T-T_{0}$ (see Fig. \ref{fig:phi_vs_T}). Among
other features, our  $\varphi_0$ vs. $T-T_{0}$ data maintains a 
continuous bifurcation
compatible with  $\varphi_0 \propto |T-T_{0}|^{1/2}$, consistent with 
previous measurements by other groups. 
Our estimations for the dependence of the anchoring angle
on temperature for MBBA 
are well in agreement with all the results reported in the
literature, both at the onset of the bifurcation 
\cite{Sonin95} and at much lower temperatures \cite{Bouchiat71}, 
extending almost $10^{\circ}$C below the transition. We 
observe a homeotropic anchoring above $T_{0}$ and a steadily increasing
anchoring angle whose saturation we are unable to observe as we enter
a regime where the disclination line collapses. We see that, provided the
anchoring angle is large enough ($\varphi_0 \geq 25^{\circ}$) a configuration
without a disclination line is possible.
On the other hand, our estimates for I52 depart from the
homeotropic anchoring configuration for $T$ above $T_{0}$, with
values for $\varphi_0$ that are about 1.7 times lower than what was
reported by Sonin {\sl et al.} \cite{Sonin95}. For  $T$ well above $T_{0}$,
we find a stable configuration with $\varphi_0 \simeq 16^{\circ}$ tilted
from the homeotropic orientation. The limited available information on 
the physical properties of I52 may account, at least partially, for these
discrepancies, even though our only assumptions are strong
anchoring and one-constant approximation for the elastic energy.
\subsection{Collapse of the disclination line in MBBA: a first-order 
transition}
%
%
%
%
\begin{figure}[t]
\leavevmode
\vbox{
\epsfxsize = \the\hsize
\epsffile{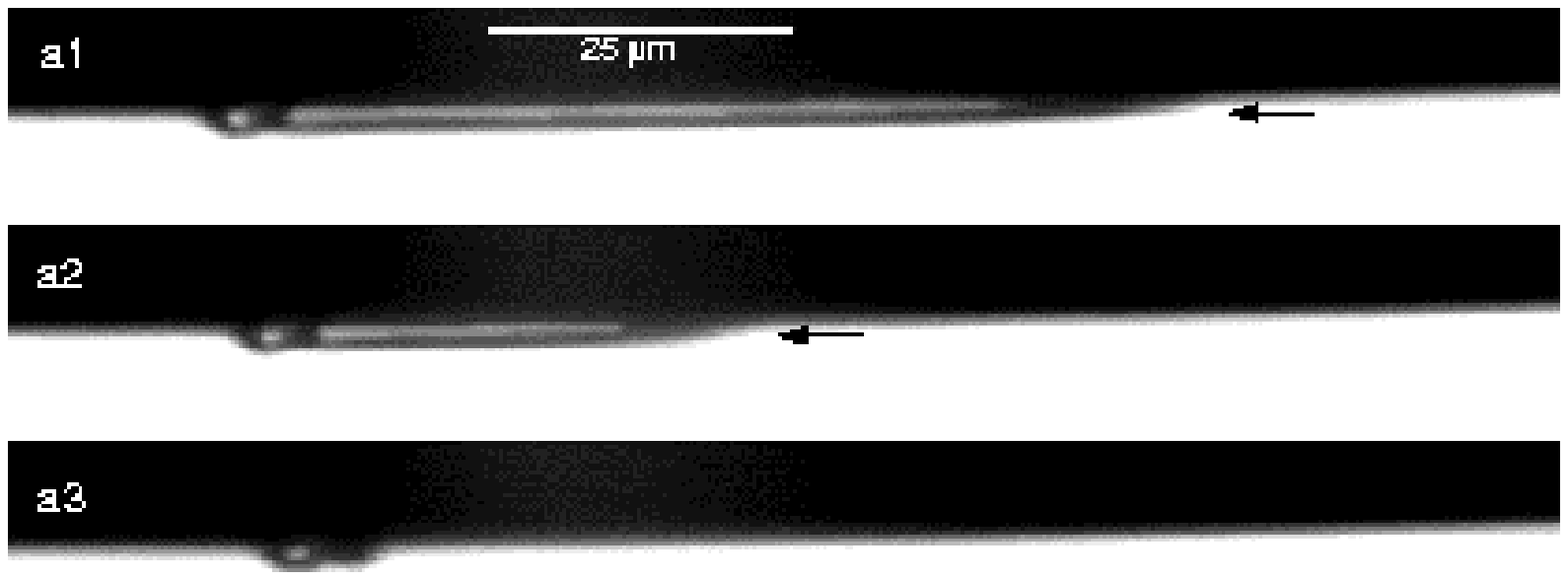}
}
\vskip 0.5cm
\null
\vbox{
\epsfxsize = \the\hsize
\epsffile{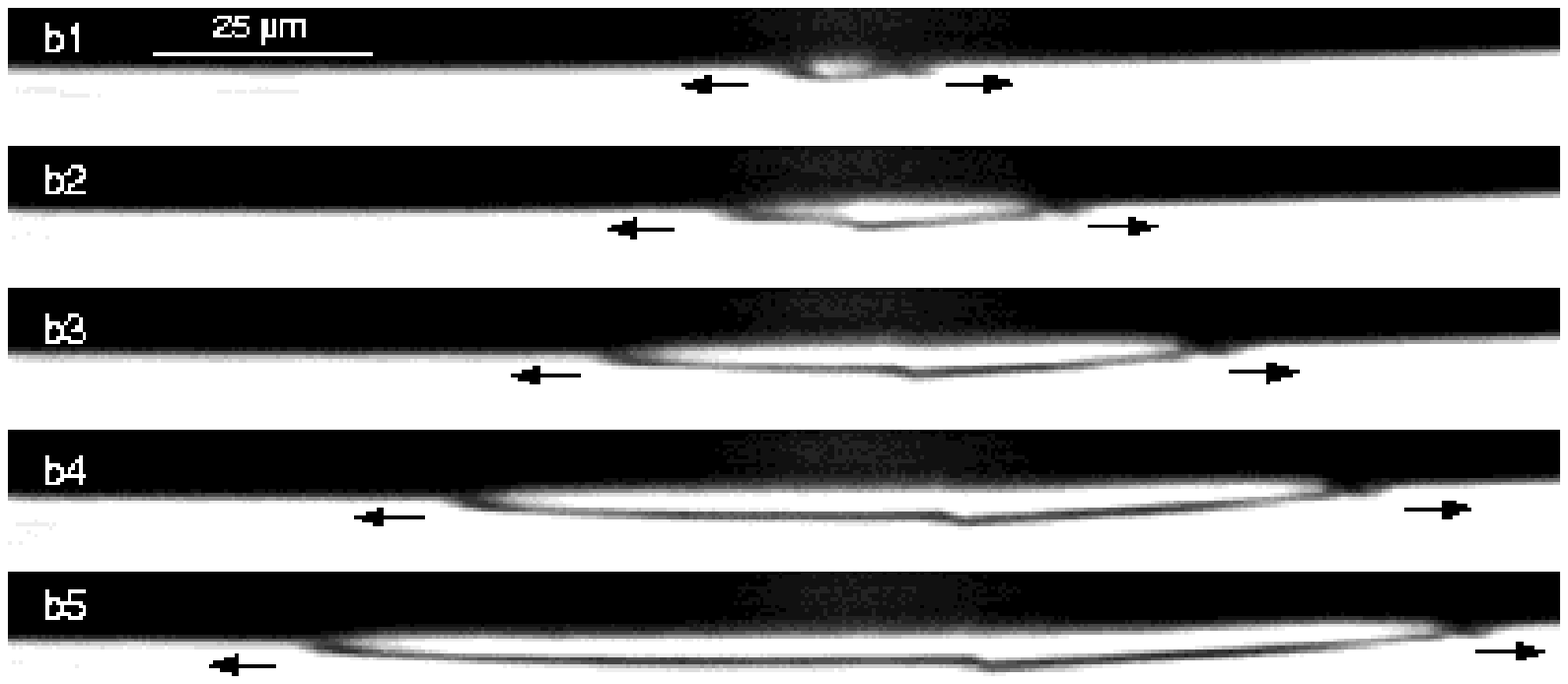}
}
\caption{
Hysteresis observed in the nucleation and in the collapse of a 
disclination line near the N/air interface 
for a $b = 220 \; \mu$m MBBA sample. The cycle is illustrated in the
diagram in Fig. \protect{\ref{fig:d_over_b_vs_T}}.
Top (a1-a3): collapse of the disclination line after a temperature quench of 
a few tenths of a degree into a final temperature of 
$T \simeq 16.3 ^{\circ}$C. 
The configuration containing a disclination line becomes unstable, and regions
without a disclination line nucleate and propagate. 
The collapsing disclination line
is pinned on a dust particle. Time sequence:  $t = 0$ (a1), $t = 8\,$s (a2),
$t = 16\,$s (a3).
Bottom (b1-b5): nucleation of the disclination line observed upon heating of 
the same sample, once the disclination line had totally collapsed. 
The disclination line starts to
regenerate around a dust particle at $T \simeq 19.3 ^{\circ}$C
and it propagates while $T$ is held constant, filling the whole
interface. The pinning
defect on the disclination line marks where the dust particle is. 
The time sequence is as follows:   $t = 0$ (b1), $t = 48\,$s (b2),
 $t = 96\,$s (b3), $t = 144\,$s (b4), $t = 192\,$s (b5). 
\label{fig:hysteresys_line}
}
\end{figure}
When the temperature is lowered below a thickness-dependent value 
(typically 10 degrees below $T_{0}$),
regions where the disclination line has collapsed start to
nucleate around dust particles on the meniscus 
(see Fig. \ref{fig:hysteresys_line}). 
These collapsed regions 
propagate until they eventually extend along all the interface, 
and the equilibrium configuration has no disclination line. Before
collapsing, the disclination line in the now metastable regions keeps 
approaching the meniscus as the temperature is lowered even further.
We have already explained that, as long as the disclination line does 
not collapse, there
is no hysteresis between heating and cooling cycles. Once the line
has collapsed, however, heating the sample does not regenerate 
the line continuously.
Instead, there is a heterogeneous nucleation process similar to the one
that leads to the 
collapse of the line upon cooling, and the temperature at which the line
is regenerated is well above the temperature at which the line started to
collapse. The diagram in Fig. \ref{fig:d_over_b_vs_T} illustrates this
hysteresis cycle.  

This dynamics suggests the coexistence of two possible configurations in the
structure of the director field for MBBA, one presenting a disclination
line and another with a continuous distortion of the director field. The
observed nucleation mechanism for the collapse and regeneration of the 
disclination line shows that the transition between these two configuration
is energy activated. A detailed study of this process would yield a
good quantitative characterization of the energy associated with the core
of the disclination line, which has yet to be done.

\section{Summary and Conclusions}
\label{sec:summary}
In this article, we have studied the nucleation and the physical properties
of a -1/2 wedge disclination line which forms near the open edge of a
nematic liquid crystal sample confined between two glass plates that induce
a homeotropic orientation of the nematic phase. The position of the
disclination line, in particular its distance to the edge of the N/air
meniscus is directly measurable, and we have related it to the physical
properties of the liquid crystal with the help of a planar model for
the structure of the director field in the presence of the N/air interface.
Since the three-dimensional nature of the structure of the director field in 
the vicinity of the disclination line is experimentally evidenced,
we have used a three-dimensional model to obtain a more accurate description
of this structure. Indeed, we have observed a significant {\sl escaped} 
component of the director field, even though the position of the disclination
line that minimizes the total energy agrees with the predictions of the planar
model, which we have therefore used since it allows to implement realistic
boundary conditions in a simple way.
We have performed different attempts at altering the position of the
disclination line by changing the anchoring parameter $Wb/K$. 
For large values of this parameter, the position of the disclination
line is completely determined by the geometry of the problem, and therefore
$d \propto b$. By changing the sample thickness, we have tried to reduce
the magnitude of $Wb/K$  so that the line aproaches the meniscus faster
than the linear dependence on $b$.  Our attempts at observing this regime 
and extracting a value for $W$
this way  have failed, most likely due to the experimental limitation in
the smallest values for $b$ we have been able to use. Under usual experimental
conditions, the disclination line will form and its position will agree
with the predictions from the {\sl infinite} anchoring approximation. 
Altering $W$ in a way that can affect the position of the disclination line
has proven to be a difficult task. Given the fact that $W$ is an intrinsic
property of the interface, one would need to consider different isotropic
media or even chemically altering the anchoring. We have been unable to
alter the nature of the anchoring strength in a stable and reproducible way. 
Another promising way to control the magnitude of $Wb/K$ is approaching
the Nematic/Smectic-A phase transition temperature. The bend elastic 
constant, $K_3$, can be made arbitrarily large by approaching that temperature.
By placing our samples in a temperature gradient, we have been able to observe
the collapse of the disclination line, which we see to start approaching the
N/air meniscus when the temperature is less than a few hundred millikelvin
above $T_{NA}$. From this experiment, we estimated $W$ in 8CB near $T_{NA}$.    
Nevertheless, this method has important experimental limitations (presence 
of a vertical temperature gradient in the samples, existence of a zig-zag
instability 
of the disclination line at very small temperature gradient, bad knowledge
of the elastic constants near the transition, etc.). 
Finally, our calculations show that a departure of the 
anchoring condition at the N/air meniscus from the homeotropic orientation
will significantly alter the position of the disclination line. Indeed,
we have been able to observe the disclination line to approach the meniscus
when materials known to have a temperature-induced anchoring angle transition
are used. Moreover, we have shown that a configuration without a 
disclination line is obtained for values of the anchoring angle above 
$25^{\circ}$ (a value that depends slightly on the material properties),
even though  energy considerations alone would suggest the possibility of
the formation of a disclination line ($Wb/K$ is large). We believe
this is the situation we encounter in the vicinity of a nematic/isotropic
liquid crystal front in directional growth experiments. Even though we
are in a regime of infinite anchoring (large $Wb/K$), a large value of the
anchoring angle (about $45^{\circ}$) guarantees that the disclination line 
will be virtual in those cases.

\acknowledgments
This work was supported by the European Research Network Contract 
No.  FMRX-CT96-0085 and the Barrande Contract No. 98010.



\begin{thebibliography}{10}

\bibitem{Frank58}
F. Frank, Disc. Faraday Soc. {\bf 25},  19  (1958).

\bibitem{Dzyaloshinskii73}
I. Dzyaloshinskii, Sov. Phys. J.E.T.P. {\bf 36},  774  (1973).

\bibitem{Candau73}
S. Candau, P. Le~Roy, and F. Debeauvais, Mol. Cryst. Liq. Cryst. {\bf 23},  
283 (1973).

\bibitem{Volovik83}
G. Volovik and O. Lavrentovich, Sov. Phys. JETP {\bf 58 (6)},  1159  (1983).

\bibitem{Ryschenkow76}
G. Ryschenkow and M.~J. Kl\'eman, J. Chem. Phys. {\bf 64}  (1976).

\bibitem{Drzaic88}
P. Drzaic, Mol. Cryst. Liq. Cryst. {\bf 154},  289  (1988).

\bibitem{Doane90}
J.~W. Doane,  in {\em Liquid Crystals: Applications and uses}, edited by B.
  Bahadar (World Scientific, Singapore, 1990), Chap.~14.

\bibitem{Cladis72}
P. Cladis and M.~J. Kl\'eman, J. Phys. France {\bf 33},  591  (1972).

\bibitem{Mihailovic88}
M. Miha\"{\i}lovic and P. Oswald, J. Phys. France {\bf 49},  1467  (1988).

\bibitem{Poulin97}
P. Poulin, H. Stark, T.~C. Lubensky, and D.~A. Weitz, Science {\bf 275},  
1770 (1997).

\bibitem{Oswald87}
P. Oswald, J. Bechhoefer, and A. Libchaber, Phys. Rev. Lett. {\bf 58},  2318
  (1987).

\bibitem{Bechhoefer89}
J. Bechhoefer, A.~J. Simon, A. Libchaber, and P. Oswald, Phys. Rev. A 
{\bf 40}, 2042  (1989).

\bibitem{Figueiredo93}
J. M. A. Figueiredo, M. B. L. Santos, L. O. Ladeira, O. N. Mesquita, Phys.
Rev. Lett. 
{\bf 71}, 4397  (1993).

\bibitem{Figueiredo96}
J. M. A. Figueiredo, O. N. Mesquita, Phys. Rev. E {\bf53}, 2423  (1996).

\bibitem{Misbah95}
C. Misbah, A. Valance, Phys. Rev. E {\bf 51}, 1282  (1995).

\bibitem{Bechhoefer95}
J. Bechhoefer, S. A. Langer, Phys. Rev. E {\bf 51}, 2356  (1995).

\bibitem{Cognard81}
J. Cognard, Mol. Cryst. Liq. Cryst. {\bf 78 supl. 1},  1  (1981).

\bibitem{Lavrentovich98}
O.~D. Lavrentovich, Liquid Crystals {\bf 24},  117  (1998).

\bibitem{Madhusudana82}
N.~V. Madhusudana and R. Pratibha, Mol. Cryst. Liq. Cryst. {\bf 89},  249
  (1982).

\bibitem{Davidov79}
D. Davidov {\it et~al.}, Phys. Rev. B {\bf 19},  1657  (1979).

\bibitem{Garland94}
C.~W. Garland and G. Nounesis, Phys. Rev. E {\bf 49},  2964  (1994).

\bibitem{Neyring71}
J. Neyring and A. Saupe, J. Chem. Phys. {\bf 54},  337  (1971).

\bibitem{Yokoyama97}
H. Yokoyama, Phys. Rev. E {\bf 57},  2938  (1997).

\bibitem{Faetti84}
S. Faetti and V. Palleschi, Phys. Rev. A {\bf 30},  3241  (1984).

\bibitem{Baudry98}
J. Baudry, S. Pirkl, and P. Oswald, Phys. Rev. E {\bf 57},  3038  (1998).

\bibitem{DeGennes}
P. G. de~Gennes and J. Prost, {\em The Physics of Liquid Crystals}, 
2nd ed. (Cambridge University Press, Cambridge, MA, 1982).

\bibitem{Bouchiat71}
M.~A. Bouchiat and D. Langevin-Crouchon, Phys. Lett. {\bf 34A},  331  (1971).

\bibitem{Sonin95}
A.~A. Sonin, A. Yethiraj, J. Bechhoefer, and B.~J. Frisken, Phys. Rev. E 
{\bf 52},  6260  (1995).



\end{thebibliography}
\end{document}